\documentclass[
    aps,
    pra,
    reprint,
    amsmath,
    amssymb,
    superscriptaddress
]{revtex4-2}

\usepackage{graphicx}
\usepackage{multirow}
\usepackage{amsmath,amssymb,amsfonts}
\usepackage{amsthm}
\usepackage{mathrsfs}
\usepackage[title]{appendix}
\usepackage{xcolor}
\usepackage{textcomp}
\usepackage{booktabs}
\usepackage{array}

\newcolumntype{C}[1]{>{\centering\arraybackslash}p{#1}}
\begin{document}

\title{Performance Comparison of Neural Network Approaches for Electric Potential Field Generation in Surface Ion Traps}

\author{Sergey Denisov}
\affiliation{P.N. Lebedev Physical Institute, Russian Academy of Sciences, Moscow, Russia}

\author{Ilya Gerasin}
\email{igerasin@lebedev.ru}
\affiliation{P.N. Lebedev Physical Institute, Russian Academy of Sciences, Moscow, Russia}

\author{Ilya Semerikov}
\affiliation{P.N. Lebedev Physical Institute, Russian Academy of Sciences, Moscow, Russia}

\begin{abstract}Quantum computing is now one of the most promising areas in modern computational science. Trapped ions are considered one of the best candidates for qubits because they have long coherence times and allow high-fidelity operations. Surface electrode ion traps are especially important because they support scalable quantum charge-coupled device architectures for advanced quantum computations. However, common simulation methods for trap design, such as the finite element method, are computationally expensive and slow down prototyping and optimization. In this paper, we compare three neural network architectures for fast electric potential field generation in surface ion traps: a Convolutional Neural Network, a Fourier Neural Operator, and Fully-Connected Layers. Our results show that the Fourier Neural Operator gives the best overall performance. It reaches an average effective potential field error of only 0.02\% and is about 10 times faster than conventional methods. These results suggest that the Fourier Neural Operator is a strong candidate for trap design and ion dynamics simulations.\end{abstract}


\maketitle

\section{Introduction}
\label{sec:introduction}
Quantum computing is one of the most promising directions in modern computational technology because it uses quantum mechanical effects to solve problems that are difficult for classical computers \cite{nielsen2010quantum}. Among the many physical platforms, trapped ions are widely seen as one of the strongest options. They offer long coherence times \cite{haffner2008quantum}, high-fidelity single-qubit \cite{gaebler2016high} and two-qubit gate operations \cite{ballance2016high}, and a high level of uniformity. Surface electrode ion traps \cite{seidelin2006microfabricated} have become an important scalable platform. They are made with microfabrication methods on substrates such as fused silica or silicon, usually with gold electrodes. Their feature sizes are typically from a few to several tens of micrometers, and ions are trapped at heights from tens to a few hundred micrometers above the surface. These microfabricated chips make it possible to build quantum charge-coupled device (QCCD) architectures \cite{kielpinski2002architecture}, where ions can be moved between different processing zones in a way similar to classical CCD devices.

Accurate calculation of electric potential fields is important for many parts of ion trap operation. These include trap design and optimization \cite{hong2016guidelines, gerasin2024optimized, abbasov2023surface, zhang2022optimization}, simulation of ion motion and confinement \cite{leibfried2003quantum, moehring2011design, singer2010colloquium}, micromotion compensation \cite{berkeland1998minimization}, and the study of trap-induced frequency shifts \cite{kim2010surface}. Fast field calculation also makes it easier to explore many design parameters and to correct experimental imperfections, such as stray electric fields, by changing voltages on compensation electrodes\cite{shu2014heating, charles2012controlling}.

Traditional design methods for surface ion traps depend on numerical simulations to calculate electric potential fields. The finite element method (FEM) divides the space into small elements and solves Laplace's equation $\nabla^2\phi = 0$ with boundary conditions. This method is accurate, but it needs a lot of computational resources. The finite difference method (FDM) places the differential operators on a grid, but it is less effective for complicated geometries. The boundary element method (BEM) reduces the problem to the boundaries, which lowers the dimensionality, but it leads to dense matrices. The solid angle method \cite{wesenberg2008electrostatics} is an efficient boundary-based approach for potential calculation in electrode structures. It works by calculating the solid angle of each electrode at every evaluation point. For a typical 2D evaluation grid of size $N \times N$, the number of evaluation points scales as $n = N^2$, while the number of boundary elements or electrodes scales as $m \sim N$. FEM requires solving a sparse linear system, and its complexity can range from $O(n)$ to $O(n^{1.5})$ depending on the solver \cite{george1973nested,briggs2000multigrid}. BEM produces a dense linear system of size $m \times m$, which gives a nominal complexity of $O(m^3)$ for direct solvers, although faster algorithms also exist \cite{greengard1997fast}. In contrast, the solid angle method calculates the potential at each point as a direct sum over all electrodes. For a fixed and small number of electrodes, this gives a complexity of $O(nm)$. One of its main advantages is that it handles singularities at electrode edges analytically, without the need for special mesh refinement \cite{wesenberg2008electrostatics}. This makes it especially suitable for our application.

Recent progress in machine learning and neural operators has created new opportunities for speeding up scientific computations \cite{li2020fourier}. Neural networks have already been used successfully in many physical simulations, including fluid dynamics \cite{kochkov2021machine}, electromagnetics \cite{khan2019deep}, and quantum systems \cite{carleo2017solving}. For example, Physics-Informed Neural Networks (PINNs) \cite{raissi2019physics} include physical laws directly in the training process, while neural operators such as Fourier Neural Operators (FNOs) \cite{li2020fourier} learn mappings between function spaces. To the best of our knowledge, there has not yet been a systematic study of neural network methods for electric field approximation in ion traps.

In this work, we compare three neural network architectures for electric potential field generation in surface ion traps. These are Convolutional Neural Networks (CNNs), which perform direct spatial field mapping using ideas from image-to-image translation; Fourier Neural Operators (FNOs), which use spectral transformations to represent global field behavior; and Fully-Connected Layers (FCL), which predict field values point by point from spatial coordinates.

We compare these approaches with classical numerical simulations and evaluate both their reconstruction accuracy and computational performance on CPU and GPU platforms, including memory usage.

\section{Methodology}

\subsection{Numerical Simulation Baseline}

We employ the solid-angle method (SAM)~\cite{wesenberg2008electrostatics} for two main purposes: (a) to generate reference data used for training and validating the neural network approaches, and (b) to serve as a classical baseline for comparison. This method provides an efficient boundary-based technique for computing electrostatic potentials in electrode geometries. It is fast and allows not to cover with the mesh all domains. For a given electrode configuration, the potential $\phi$ at a point $\mathbf{r}$ can be expressed as
\begin{equation}
\phi(\mathbf{r}) = \frac{1}{4\pi\epsilon_0} \sum_i V_i \Omega_i(\mathbf{r})
\end{equation}
where $V_i$ is the voltage of electrode $i$, and $\Omega_i(\mathbf{r})$ is the solid angle subtended by electrode $i$ at point $\mathbf{r}$. For $n$ evaluation points and $m$ electrodes, this method gives accurate potential values with computational complexity $O(nm)$ and handles singularities at electrode edges well.

\subsection{FEM baseline}
To benchmark in terms of speed our approaches against a standard out-of-the-box method, we compare our calculations with COMSOL FEM simulations. We employ the 2D Electrostatics module to compute the potential generated by a five-rail surface trap. The electrodes are modeled as 1~$\mu$m-thick structures placed within a vacuum box that spans the full width of the electrodes and has a height of 200~$\mu$m. A surface layer representing an Infinite Element Domain with a thickness of 50~$\mu$m is added. The region of interest is discretized with a $50 \times 50$ mesh, matching the resolution used in the neural network approach, while the surrounding domain is meshed using an extremely coarse grid. The stationary solver selected for this problem is MUMPS, using the default settings recommended by COMSOL. For performance evaluation, we perform a parametric sweep over the set of widths for the central DC and RF electrodes and compute the average time per simulation run. This scenario closely mimics the optimization process we follow when designing traps in our laboratory.

\subsection{Neural Network Architectures}

\subsubsection{Convolutional Neural Network (CNN)}
\begin{figure}[htbp]
 \centering
        \includegraphics[width=0.98\linewidth]{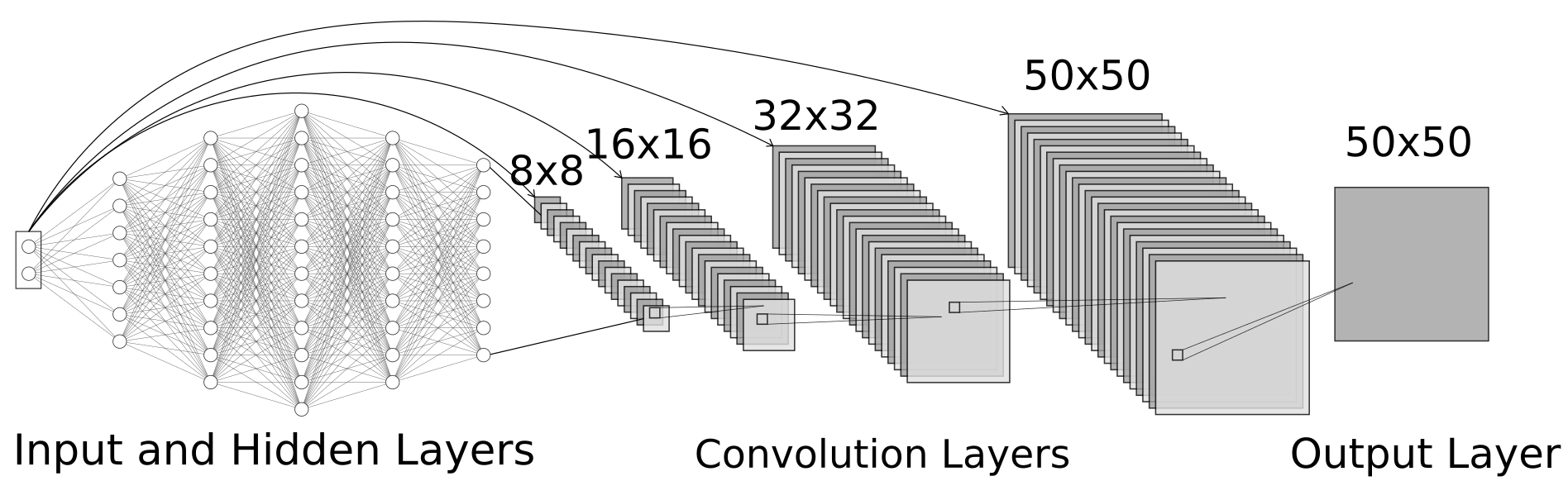}
\caption{CNN architecture diagram showing encoder-decoder structure with progressive upsampling from 8$\times$8 to 50$\times$50 resolution. The network takes 2 input parameters and outputs a complete 50$\times$50 potential field. This image was made with the help of Alexander Lenail's tool NN-SVG.}
\label{fig:cnn_architecture}
\end{figure}

The CNN architecture is inspired by image-to-image translation networks \cite{isola2017image}. It uses a multi-resolution approach. The network takes two input parameters, namely the RF electrode widths and the separation distance, and produces a full $50\times50$ potential field (see Fig. \ref{fig:cnn_architecture}). The process includes several steps: first, the two electrode parameters are used as input; next, fully-connected layers map them into a latent space; then convolutional layers with bilinear interpolation gradually upsample the representation from $8\times8$ to $50\times50$; finally, the network outputs the full $50\times50$ potential field. The model has about 15 million parameters. To keep geometric information through the network, the input parameters are added at different resolutions using skip connections.

\subsubsection{Fourier Neural Operator (FNO)}
\begin{figure}[htbp]
\centering
\includegraphics[width=0.98\linewidth]{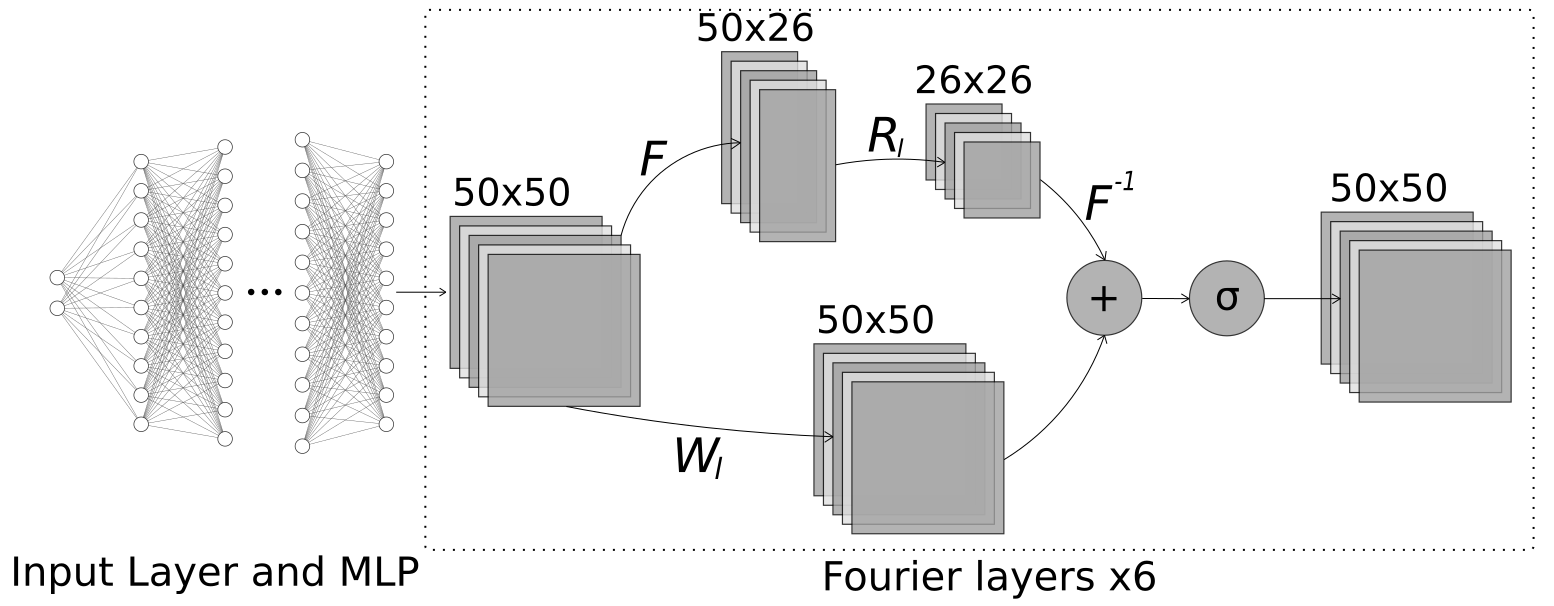}
\caption{FNO architecture diagram showing Fourier layers with spectral transformations. The network processes inputs through an encoder, 6 Fourier layers with frequency truncation, and a projection layer to generate 50$\times$50 output fields. This image was made with the help of  Alexander Lenail's tool NN-SVG.}
\label{fig:fno_architecture}
\end{figure}

The FNO architecture \cite{li2020fourier} uses spectral transformations to represent global field behavior. It is especially suitable for smooth physical fields such as electric potentials. A Multi-Layer Perceptron (MLP), which is a fully-connected neural network with several hidden layers, is used as an encoder to map the input parameters into a spatial representation (see Fig. \ref{fig:fno_architecture}). Then, six Fourier layers perform spectral convolutions with mode truncation to $(26,26)$ modes. After that, a projection layer with $1\times1$ convolutions reduces the channel dimension from 64 to 1 and produces a $50\times50$ potential field. The full network has about 8 million parameters.

The Fourier layers apply the following operation:
\begin{equation}
z_{l+1} = \sigma\left(\mathcal{F}^{-1}\left(R_l \cdot \mathcal{F}(z_l)\right) + W_l z_l\right)
\end{equation}
where $z_l$ is the input to layer $l$, $\mathcal{F}$ and $\mathcal{F}^{-1}$ are the forward and inverse Fourier transforms, $R_l$ is the complex spectral weight tensor acting on the truncated Fourier modes, $W_l$ is a local linear transformation implemented by convolution, and $\sigma$ is the activation function. This structure allows the model to learn global operators efficiently while still keeping local spatial information through the residual connection.

\subsubsection{Fully-Connected Layers (FCL)}
\begin{figure}[htbp]
 \centering
        \includegraphics[width=0.98\linewidth]{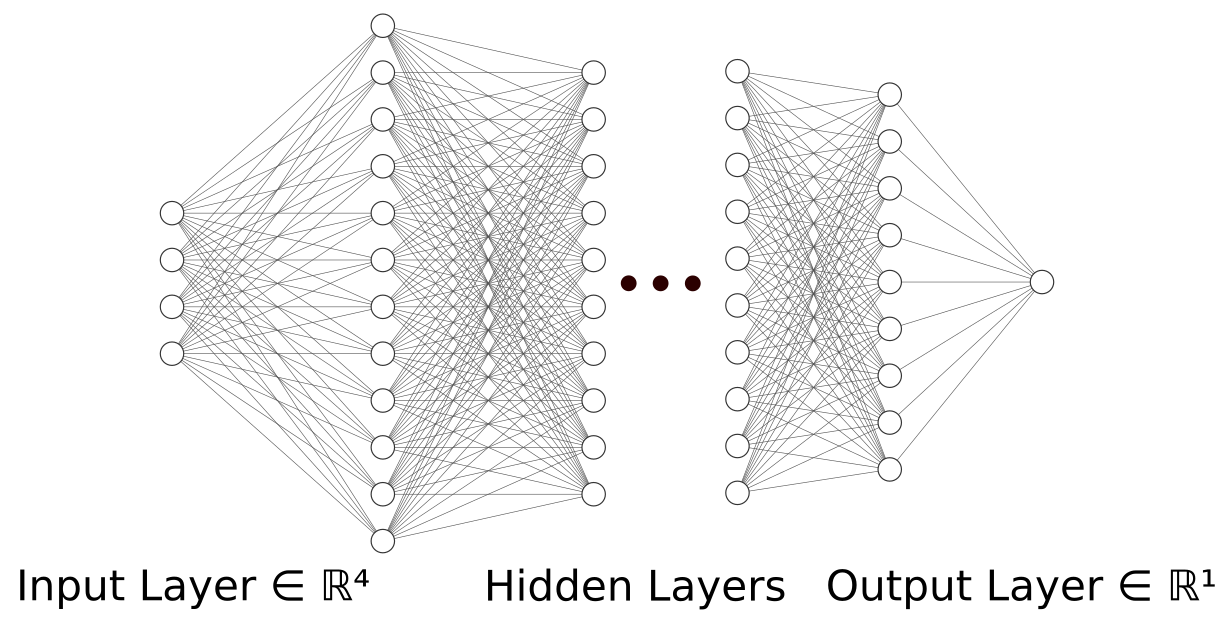}
 \caption{FCL architecture diagram showing three specialized networks for potential, field magnitude (in log scale), and field sign prediction. The system takes both electrode parameters and spatial coordinates as inputs. This image was made with the help of Alexander Lenail's tool NN-SVG.}
\label{fig:fcl_architecture}
\end{figure}

The FCL approach uses networks that take spatial coordinates as additional inputs, which allows continuous field representation. The system includes three specialized networks: a 12-layer network that predicts electric potential and has about 4 million parameters, a 13-layer network that predicts $\lg(|\mathbf{E}|)$ and has about 5 million parameters, and a 6-layer network that predicts the sign of the electric field and has about 0.5 million parameters (see Fig. \ref{fig:fcl_architecture}).

The electric field is reconstructed as
\begin{equation}
\mathbf{E} = \text{sign} \times 10^{\lg(|\mathbf{E}|)}
\end{equation}

We predict $\lg(|\mathbf{E}|)$ instead of the direct field value because electric fields in ion traps can vary over several orders of magnitude. This makes direct regression difficult for neural networks. The logarithmic transformation reduces this large dynamic range and makes the learning problem easier.

\subsection{Training Methodology}

All models were trained using Mean Squared Error (MSE) loss:
\begin{equation}
\mathcal{L} = \frac{1}{N} \sum_{i=1}^N (y_i - \hat{y}_i)^2
\end{equation}
where $y_i$ is the ground truth value taken from output of SAM and $\hat{y}_i$ is the prediction made by the neural network.

We also tested a masking strategy and compared masked and unmasked models. Points with pseudopotential magnitude above the threshold $\epsilon = 0.3 \text{ }eV$ were excluded from the loss calculation. This threshold corresponds to regions where there is no equilibrium and where useful trap parameters cannot be extracted. As a result, the training focused on the physically relevant region for ion confinement and dynamics, within a radius of about $10 \text{ }\mu m$ around the equilibrium point. This improved the accuracy of parameter extraction.

We also explored other approaches, including Physics-Informed Neural Networks (PINNs) with Laplacian constraints ($\nabla^2\phi = 0$) and relative error metrics, but these performed worse than the good old fashion MSE with masking. There are several reasons for this. First, because our simulations use a finite grid resolution, the discrete Laplacian does not become exactly zero. Instead, it approaches a small constant value that depends on grid spacing, which introduces numerical error into the physics constraint. Second, adding physics-based loss terms such as $\nabla^2\phi = 0$ makes the optimization problem more difficult, because the network must fit the data and satisfy the physical constraint at the same time. In practice, this multi-objective optimization was less stable and required careful balancing of the loss weights.

\subsection{Trap Parameter Extraction}

We extracted several relevant ion trap parameters from the generated fields. These include the equilibrium height ($h$), defined as the position of minimum effective potential; the secular frequency (freq), obtained from the second derivatives of the potential at the equilibrium point; the trap depth, defined as the maximum confining potential energy; and the stability parameter ($q$), taken as the Mathieu parameter derived from the field curvature and related to the ratio of the secular frequency to the RF frequency.

These parameters are essential for predicting ion behavior and overall trap performance in quantum operations.

\section{Experimental Results}

\subsection{Field Reconstruction Accuracy}

Table \ref{tab:field_accuracy} summarizes field reconstruction performance. Success rate indicates the percentage of generated fields that produced physically meaningful results. Pushing the limits of these neural networks, they would sometimes yeild generated fields of poor quality and/or fields that where completely wrong. These fields where not considered for the accuracy comparison results and later the trap parameter error results. The criteria of a "meaningful" result was taken as: the error of all the extracted parameters did not exceed 100\%.
\begin{table*}[t]
\centering
\caption{Field Reconstruction Accuracy Comparison}
\begin{tabular}{l c c c}
\hline

Model&Potential Error (\%) &Field Error (\%)&Success Rate (\%)\\

\hline
FNO (masked) & 0.02 & 0.15 & 100 \\
FNO (unmasked) & 0.03 & 0.22 & 100 \\
CNN (masked) & 0.49 & 3.81 & 100 \\ 
CNN (unmasked) & 1.30 & 9.45 & 90 \\
FCL & 0.40 & 2.00 & 100 \\
\hline
\end{tabular}
\label{tab:field_accuracy}
\end{table*}

The FNO gives the best overall performance, with an average error of only 0.02\% in effective potential field reconstruction. For CNN and FNO, field values were obtained by numerically differentiating the predicted potential fields, while FCL directly predicts the field.

\subsection{Trap Parameter Extraction}

Table \ref{tab:parameter_accuracy} shows the accuracy of the extracted trap parameters, which are critical for real trap operation.

\begin{table}[htbp]
\centering
\caption{Trap Parameter Extraction Error (\%)}
\begin{tabular}{l c c c c}
\hline
Model & h & q & freq & depth \\
\hline
FNO (masked) & 0.1 & 1.0 & 4.1 & 2.8 \\
FNO (unmasked) & 0.5 & 3.1 & 10.0 & 3.1 \\
CNN (masked) & 0.9 & 4.5 & 23.8 & 15.6 \\
CNN (unmasked) & 1.4 & 10.2 & 39.9 & 33.1 \\
FCL (from E-field) & 4.0 & 11.0 & 9.0 & 10.0 \\
FCL (from $\phi$) & 0.6 & 1.9 & 3.3 & 4.7 \\
\hline
\end{tabular}
\label{tab:parameter_accuracy}
\end{table}

The masked FNO achieves very strong accuracy in parameter extraction. The FCL approach gives better frequency accuracy when the parameter is extracted from the potential rather than directly from the predicted electric field.

\subsection{Computational Performance}
Table \ref{tab:computational_performance} compares the inference times of the SAM, COMSOL (we use default settings for multithreading) and the neural models on both CPU (AMD Ryzen 9 7950X, 16 cores) and GPU platforms (NVIDIA GeForce RTX 4060). The reported times are averages over 1000 inference runs.

\begin{table}[htbp]
\centering
\caption{Computational Performance Comparison}
\begin{tabular}{l c c}
\hline
Method & GPU Time & CPU Time \\
\hline
SAM & - & 7.8 ms  \\
COMSOL & - & 980 ms \\
FNO & 0.53 ms & 5.7 ms \\ 
CNN & 2.4 ms & 17 ms  \\
FCL (full field) & 62 ms & 230 ms\\
FCL (per point) & 25 $\mu$s & 91 $\mu$s \\
\hline
\end{tabular}
\label{tab:computational_performance}
\end{table}

All neural network approaches provide clear speed improvements on GPU hardware. For FCL, the reported time is per point. Generating a full $50\times50$ field requires 2500 separate evaluations. Although FCL is slow for full-field generation, it has the advantage of continuous evaluation at arbitrary coordinates without interpolation.

COMSOL exhibits extremely slow performance. The first reason is that meshing introduces an overhead. One needs to use boundary conditions such as the Infinite Element Domain, which are necessary for correct field calculations. In our calculations for a $50\times50$ area of interest, COMSOL requires approximately 4,000 elements. The second reason is that the meshing process is included in the evaluation time. Unfortunately, this workflow cannot be avoided—this is the price when using out-of-the-box tools.

\section{Discussion}

The strong performance of the FNO comes from its ability to capture global field relationships through spectral transformations. Electric potential fields in ion traps are naturally smooth and satisfy Laplace's equation, which makes them well suited for spectral representation. The Fourier layers capture these smooth global variations effectively, while local convolutional parts help handle edge effects near the electrodes.

The CNN also performs reasonably well, but it has more difficulty with long-range dependencies. This becomes especially visible in parameters related to dynamics, such as oscillation frequency (23.8\% error) and trap depth (15.6\% error), because these depend on accurate estimation of field curvature. The progressive upsampling process can introduce artifacts that reduce accuracy in these cases.

The FCL approach shows a different error pattern, even though its number of parameters is similar to the other models. Point-wise evaluation naturally gives continuous output, but because there is no explicit spatial consistency constraint, the predicted fields can be noisier. At the same time, direct field prediction is useful for ion dynamics simulations because it avoids numerical differentiation when calculating forces.

The masking strategy is very important for both training stability and final accuracy. By focusing learning on regions with strong field gradients, where confinement forces act on the ions, masking improves the extraction of physically relevant parameters. For the FNO, masking reduces the depth error from 3.1\% to 2.8\% and the frequency error from 10.0\% to 4.1\%. The CNN benefits even more, and unmasked training produces physically unrealistic fields in 10\% of cases.

For fast ion trap design and optimization, the FNO offers the best balance between accuracy and speed. It provides about a 10 times speedup over the solid angle method on CPU while keeping the error below 5\% for all major parameters. This makes it possible to explore design spaces much faster than with traditional numerical methods. It is also worth noting that many standard differential equation solvers and simulation tools usually run on CPU, while neural network acceleration is especially effective on GPUs.

For ion dynamics simulations, the best choice depends on the task. The FNO is the fastest for generating complete fields, while the FCL allows continuous force evaluation at arbitrary points without interpolation artifacts. The CNN has the fastest inference time, but its lower derivative accuracy makes it less suitable for simulations where precise field curvature is important.

\section{Conclusion}

This study presents a comparison of neural network methods for electric potential field generation in surface ion traps. Our results show that neural networks can serve as a practical alternative to traditional numerical simulations, and that the Fourier Neural Operator is the most effective architecture among the models we tested.

The FNO reaches very high accuracy, with only 0.02\% field error, while also providing major computational speed advantages. This makes it a strong option for fast trap design and optimization. We also show that masking is an important part of the training process, because it focuses the model on physically relevant regions and improves the accuracy of extracted trap parameters.

In future work, it would be useful to extend this approach to 3D field generation for more complex trap geometries, include time-dependent fields for RF simulations, develop hybrid models that combine the strengths of different architectures, and apply these methods to multi-ion crystals and more complicated trapping potentials.

The demonstrated speed and accuracy of neural field generation opens new possibilities for real-time trap optimization, automated design processes, and comprehensive parameter space exploration in quantum computing applications.

\section*{Declarations}
\section*{Funding}
The authors declare that no funds, grants, or other support
were received during the preparation of this manuscript.

\bibliography{main_sn}
\end{document}